# Quantized electromagnetic tornado in pulsar vacuum gap

V.M.Kontorovich

*Institute of Radio Astronomy of National Academy of Sciences of Ukraine*
*V.N.Karazin National University, Kharkov*
*vkont1001@yahoo.com, vkont@ri.kharkov.ua*

Giant pulses (GP), sporadically observed in a small number of pulsars, are a riddle which is not solved yet. GP is characterized by enormous flux density[1], extremely small pulse duration [2], presence of circular polarization of both directions[3], power distribution by energies[4] and is mainly located in the narrow window with respect to average pulse position[5]. All these features fundamentally distinguish GPs from ordinary pulses.

Some extremely small giant pulse durations reaching several nanoseconds give us evidence to suggest that they may arise in a vacuum gap[6] in the process of primary electron acceleration[7] to the gamma-factors of order of $10^7$. Really, relativistic aberration in the primary electron beam reduces the cone angle of radiation to the values $\delta\varphi \sim 10^{-7}$, and the rotation with periods $P \approx 2\pi \cdot 10^{-2}$ s leads in this case to the nanosecond pulse durations $\delta t \approx \delta\varphi \cdot P/2\pi \sim 10^{-9}$ s. This explanation is consistent also with the fact that giant pulses are observed only in the rapidly rotating pulsars. Thus, we have to deal with the emission of an individual discharge in the vacuum gap[8]. Coulomb repulsion of particles in the puncture spark of the discharge leads to spark rotation around its axis in the crossed fields (an electromagnetic tornado), which provokes to the appearance of observed circular polarization[3] of giant pulses.

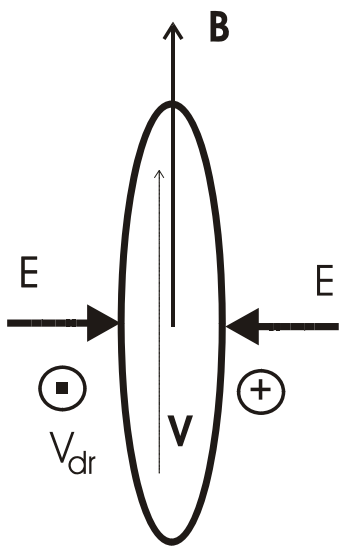

Fig.1. In the discharge bunch filament the Coulomb particle repulsion leads to rotation in the crossed electric and magnetic fields around the bunch axis transforming the bunch into electromagnetic tornado. For electrons and positrons the rotation directions are opposite. The rotation can cause the circular polarization which is observed in GPs.

In this report we have found the solution, one branch of which describes the electromagnetic tornado, and another describes a changed cyclotron rotation of the electron. Really, the infinitely thin charged filament in 3D-space causes, as known, logarithmic potential $\phi \propto \ln r$ and radial electric field $E_r \propto 1/r$, decreasing inversely proportional to distance from the charge. Accordingly, for drift velocity $V_\varphi = cE_r/H$ we have the same dependence upon the distance $V_\varphi \propto 1/r$. Thus, at large distances from the discharge axis we have a movement with constant circulation (and solid-type rotation on the small ones).

Owing to the drift the discharge channel turns into a peculiar vortex which resembles the well-known potential vortex – tornado. However, contrary to the hydrodynamic nature of the usual tornado, the vacuum gap tornado has a purely electromagnetic origin. The equations of motion of the electromagnetic tornado in the plane orthogonal to a magnetic field in complex variables $\xi = x + iy$ take the form $dw/dt - i\omega_c w = \xi \cdot eE/mr, \; w = d\xi/dt$, where $E$ is the field of a space charge, and $\omega_c$ is the cyclotron frequency. In the axial electric field $E=E(r)$ depending only on the distance $r$ from the axis, that we will believe being a constant parameter, these equations allow solutions $\xi = \xi_0(r)e^{i\Omega(r)t}$. The frequencies $\Omega$ obey the equation $\Omega^2 - \omega_c \Omega + eE/mr = 0$ which roots are equal $2\Omega_\pm = \omega_c \pm \sqrt{\omega_c^2 - 4eE/mr}$. Transition to the l-system in which the bunch moves with relativistic velocity results in replacement $\Omega \to \Omega/\Gamma$ where $\Gamma$ – is the Lorentz-factor of the bunch. In the pulsar conditions there is a small parameter $4eE/mr\omega_c^2 \ll 1$ by which expansion gives us the values $\Omega_+ \approx \omega_c - eE/mr\omega_c$ and $\Omega_- \approx cE/Hr$.

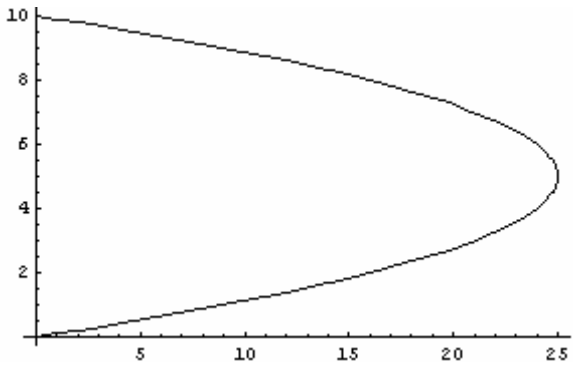

Fig.2. The angular velocities $\Omega_\pm$ dependence on quantity $eE/mr$ — function of the radial electric field $E$ formed by the discharge space charge (cf. with results for charged beams in plasma[9]).

The first root is in agreement with the common cyclotron mode slightly modified by the electric field. The second root, being of main interest to us, corresponds to the drift in the crossed fields. The solution $V_\varphi = r \cdot \Omega_-$ describes the electromagnetic tornado in the vacuum gap where the repulsion field of the space charge is compensated by the Lorentz force, the radial movement is absent, and rotation is specified by the drift in the crossed fields. Due to this rotation, the circular polarization will appear in the discharge radiation.

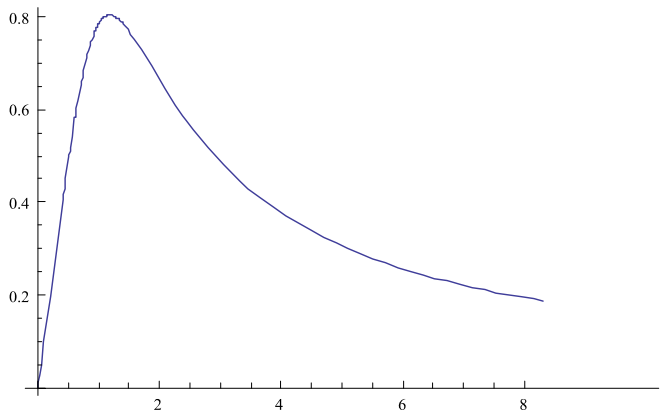

Fig.3. The dependence of the linear velocity of electromagnetic tornado $V_\varphi$ and the radial electric field $E$ from the distance from the tornado axis.

The quasi-classical quantization is possible by using the condition $mrV_\varphi = n \cdot \hbar$, which leads to the angular frequency $\Omega_n = n\hbar/mr^2$ and to the macroscopic particle linear density on the axis that does not contain any model parameters and only depends on the magnetic field: $N = n \cdot \hbar B/emc$. Here $n$ is integer or half-integer, $B$ is the magnetic field in the cavity of order of $10^{12}$G. The full Goldreigh-Julian current flowing through the gap is provided by $q$ vortex lines. In the ground state it is required for that about $q \approx 10^7$ lines. For a smaller $q$ the Rydberg states with $n \gg 1$ is required. Rotation frequencies form the bands the borders of which are determined by tornado internal and external radii. Such structure might explain the frequency bands observed in GP spectrum.

Not only quasi-classical but also exact quantum problem may be stay and solve for the electromagnetic tornado.

The electromagnetic tornado scheme in the polar vacuum gap not only realizes the origin of a circular polarization in GP but also supports the idea[*] that radio emission appears in the pulsar internal vacuum gap which is a cavity-resonator[10,11] (with respect to the radio-frequency radiation) excited by particles accelerated in a longitudinal electric field in the gap. GPs localization[4,5] near the wave-guides and slots means that the magnetosphere in the region of open field lines is basically not transparent for radiation except for the localization places. Energy emission through the breaks existing or accidentally appearing in the magnetosphere of open field lines corresponds to giant pulses. Extremely high energy density[5] of GPs of order of $10^{15}$ erg/cm$^3$ seems to be a key moment[12].

It will be noted that trying to explain the GPs by strongly nonlinear effects in magnetosphere plasma where different variants of two-stream instability are realized[13] requires considering such mechanisms as modulation instability[14], Zakharov plasma wave collapse[3] (the more popular!), reconnection of magnetic field lines[15], induced scattering in narrow beams[16] etc. (see also references in review[17]). GP phenomena resemble also those of "rogue" or "freak" waves in nonlinear hydrodynamics of sea waves (see, e. g., the conference materials[18]). Both that point of view and the above-stated may exist not interfering each other.

This text is an extended version of the thesis of the report[19] (which presentation will be placed at Landau's Institute for Theoretical Physics site http://itp.ac.ru).

---

[*] The article with a detail discussion of this idea sent now for publication.